\documentstyle[11pt,epsf]{article}
\begin{document}
\def\1{\'{\i}}

\centerline{\Large\bf Cyclotomy and Ramanujan Sums in Quantum Phase Locking}
\vspace*{1cm}

\centerline{Michel Planat$^\dagger$\footnote{e-mail: planat@lpmo.edu; Fax: 33(0)381853998} and Haret C. Rosu$^\ddagger$\footnote{e-mail: hcr@ipicyt.edu.mx;
Fax: 524448335412}}
\vspace*{2mm}

\centerline{\small\it $^\dagger$Laboratoire de Physique et M\'etrologie des Oscillateurs du CNRS, 25044 Besan\c con Cedex, France}

\centerline{\small\it $^\ddagger$Applied Mathematics, IPICyT, Apdo. Postal 3-74
Tangamanga, San Luis Potos\1, M\'exico}

\vspace*{5mm}

\begin{center}
\small
\begin{minipage}{14cm}

{\bf Abstract}.  Phase-locking governs the phase noise in classical clocks through effects described in precise 
mathematical terms. We seek here a quantum counterpart of these effects by working in a
finite Hilbert space. We use a coprimality condition to define phase-locked quantum states and the corresponding
Pegg-Barnett type phase operator. Cyclotomic symmetries in matrix elements are revealed and related
to Ramanujan sums in the theory of prime numbers. 
The employed mathematical procedures also emphasize the isomorphism between
algebraic number theory and the theory of quantum entanglement.

\bigskip


\end{minipage}
\end{center}

\normalsize

Time and phase are amongst the most federal concepts of
science. Hersh mentions Kant as putting time before number and making the intuition of 
time the origin of arithmetics \cite{hersh}.

The present work can be considered as an extension of a longstanding effort to model
phase noise and phase-locking effects that are found in highly
stable classical oscillators. It was unambiguously demonstrated that the
observed variability (i.e., the 1/f frequency noise of such
oscillators) is related to the finite dynamics of states during
the measurement process and to the precise filtering rules that
involve continued fraction expansions, prime number decomposition,
and hyperbolic geometry \cite{Planat1}-\cite{Planat3}.
We are interested here in possible quantum counterparts of these effects. The
problem of defining quantum phase operators was initiated by Dirac
in 1927 \cite{Dirac}. After more than three quarters of a century, it is still an interesting and essentially open
problem. For an excellent
review, see \cite{Reviews}. Our starting point is the Pegg and
Barnett (PB) quantum phase formalism \cite{Pegg1}-\cite{Pegg2} where the calculations
are performed in a Hilbert space $H_q$ of finite dimension $q$ \cite{vourdas}. In the PB framework the
phase states are the so-called discrete quantum Fourier transforms (QFT) of the number states, i.e., 
the following superpositions 
\begin{equation}
|\theta _p\rangle=q^{-1/2}\sum_{n=0}^{q-1}\exp\left(\frac{2i\pi p n
}{q}\right)|n\rangle~. \label{eq1}
\end{equation}
The states $|\theta_p\rangle$ form an orthonormal set and in
addition the projector over the subspace of phase states is
$\sum_{p=0}^{q-1}|\theta_p\rangle \langle\theta_p|=1_q$ where
$1_q$ is the identity operator in $H_q$. The inverse QFT implies 
$|n\rangle=q^{-1/2}\sum_{p=0}^{q-1}\exp(-\frac{2i\pi p n
}{q})|\theta _p\rangle$. As the set of number states $|n\rangle$,
the set of phase states $|\theta_p\rangle$ is a complete set
spanning $H_q$. In addition, the QFT operator is a $q\times q$
unitary matrix of entries
$\kappa_{pn}^{(q)}=\frac{1}{\sqrt {q}}\exp(2i\pi \frac{pn}{q})$.

From now on, we restrict ourselves to the class of phase states $|\theta' _p\rangle$ for which $p$ and $q$  
satisfy the coprimality condition $(p,q)=1$, where $(p,q)$ is the greatest common divisor of $p$ and $q$. 
Differently from the phase states
(\ref{eq1}), $|\theta' _p\rangle$ form an orthonormal base of a Hilbert space whose dimension is given by 
the number of irreducible fractions $p/q$. This number is known in Mathematics as the Euler
totient function $\phi(q)$.

Guided by the analogy with the classical situation \cite{Planat1} we call the irreducible states 
the {\em phase-locked quantum
states}. In the space of quantum states, they generate a cyclotomic lattice $L$ \cite{cyclot1}-\cite{cyclot5} with a
generator matrix $M$ of matrix elements $\kappa_{pn}^{'(q)}$ and
$(p,q)=1$ and size $\phi(q) \times \phi(q)$. We notice that the corresponding Gram
matrix of $L$, namely $H=M^{\dag}M$, has matrix elements
$h_{n,l}^{(q)}=c_q(n-l)$, which are Ramanujan sums
\begin{equation}
c_q(n)=\sum_{\stackrel{p=1}{(p,q)=1}}^q \exp\left(2i\pi \frac{p}{q}
n\right)=\frac{\mu(r)\phi(q)}{\phi(r)}~, \label{eq2}
\end{equation}
where $r=q/(q,n)$.
Ramanujan sums are thus defined as sums over the primitive
characters $\exp(2i\pi \frac{pn}{q})$, $(p,q)=1$, of the group
$Z_q=Z/qZ$. In the equation above, $\mu(r)$ is the M\"obius
function which is $0$ if the prime number decomposition of $r$
contains a square, $1$ if $r=1$, and $(-1)^k$ if $r$ is the
product of $k$ distinct primes \cite{Hardy}. Ramanujan sums are
relative integers that are quasiperiodic in $n$ with quasi
period $\phi(q)$ and aperiodic in $q$ with a type of
variability imposed by the M\"obius function (introduced by Ramanujan in the 
context of the Goldbach conjecture). Recently, we demonstrated that Ramanujan's sums are also useful in the context of 
signal processing as an arithmetical alternative to the discrete classical Fourier
transform \cite{Planat2}\footnote{In the discrete Fourier transform the signal processing is 
performed by using all roots of unity of the
form $\exp(2i\pi p/q )$ with $p$ from $1$ to $q$  and taking their 
$n^{\rm{th}}$ power $e_p(n)$ as basis function.
We generalized the classical Fourier analysis by using Ramanujan sums 
$c_q(n)$ as in (2) instead of $e_p(n)$.
This type of signal processing is more appropriate for arithmetical 
functions than is the ordinary discrete
Fourier transform, while still preserving the metric and orthogonal properties of the latter. 
Notable results relating arithmetical functions
to each other can be obtained using Ramanujan sums expansions while the 
discrete Fourier transform would
lead instead to low frequency tails in the power spectrum.}.

The projection operator over the subset of phase-locked quantum
states $|\theta' _p\rangle$ is calculated as the coprimality-constrained sum of one-state projectors, i.e.,
\begin{equation}
P_q^{\rm lock}=\sum_{\stackrel{p=0}{(p,q)=1}}^{q-1}P'_p=\sum_{\stackrel{p=0}{(p,q)=1}}^{q-1}|\theta'_p\rangle
\langle\theta'_p|=\frac{1}{q}\sum_{n,l=0}^{q-1} c_q(n-l)|n\rangle
\langle l|,
 \label{eq3}
\end{equation}
so that the matrix element of the projection is such that
$q\langle n|P_q|l\rangle = c_q(n-l)$. This is similar to the elements of the cyclotomic Gram matrix and sheds light on the
equivalence between cyclotomic lattices of algebraic number theory
and the quantum theory of phase-locked states. The
quantum phase-locking operator is defined
\begin{equation}
\Theta_q^{\rm lock}=\sum_{\stackrel{p=0}{(p,q)=1}}^{q-1} \theta_p
|\theta'_p\rangle \langle\theta'_p|=\pi
P_q~,~~{\rm where}~\theta _p=2\pi\frac{p}{q}~. \label{eq4}
\end{equation}
The Pegg and Barnett phase operator $\Theta_q$ \cite{Pegg1}-\cite{Pegg2} results from the removal of
the coprimality condition $(p,q)=1$. 
Using the number operator
$N_q=\sum_{n=0}^{q-1} n|n \rangle \langle n|$ a
generalization of Dirac's commutator $[\Theta_q,N_q]=-i$ is obtained. 

In contrast, the phase number commutator for phase-locked states as
calculated from (\ref{eq4}) is
\begin{equation}
C_q^{\rm lock}=[\Theta_q^{\rm lock},N_q]=\frac{\pi}{q}\sum_{n,l=0}^{q-1}(l-n)c_q(n-l)|n\rangle\langle
l|, \label{eq5}
\end{equation}
and has antisymmetric matrix elements $\langle
l|C_q^{\rm lock}|n\rangle=\frac{\pi}{q}(l-n)c_q(n-l)$. 

The finite and cyclotomic
quantum mechanical rules are mostly encoded in the expectation values of
the phase operator and phase variance (see Figures 1-3).
\vskip 1ex
\centerline{
\epsfxsize=220pt
\epsfbox{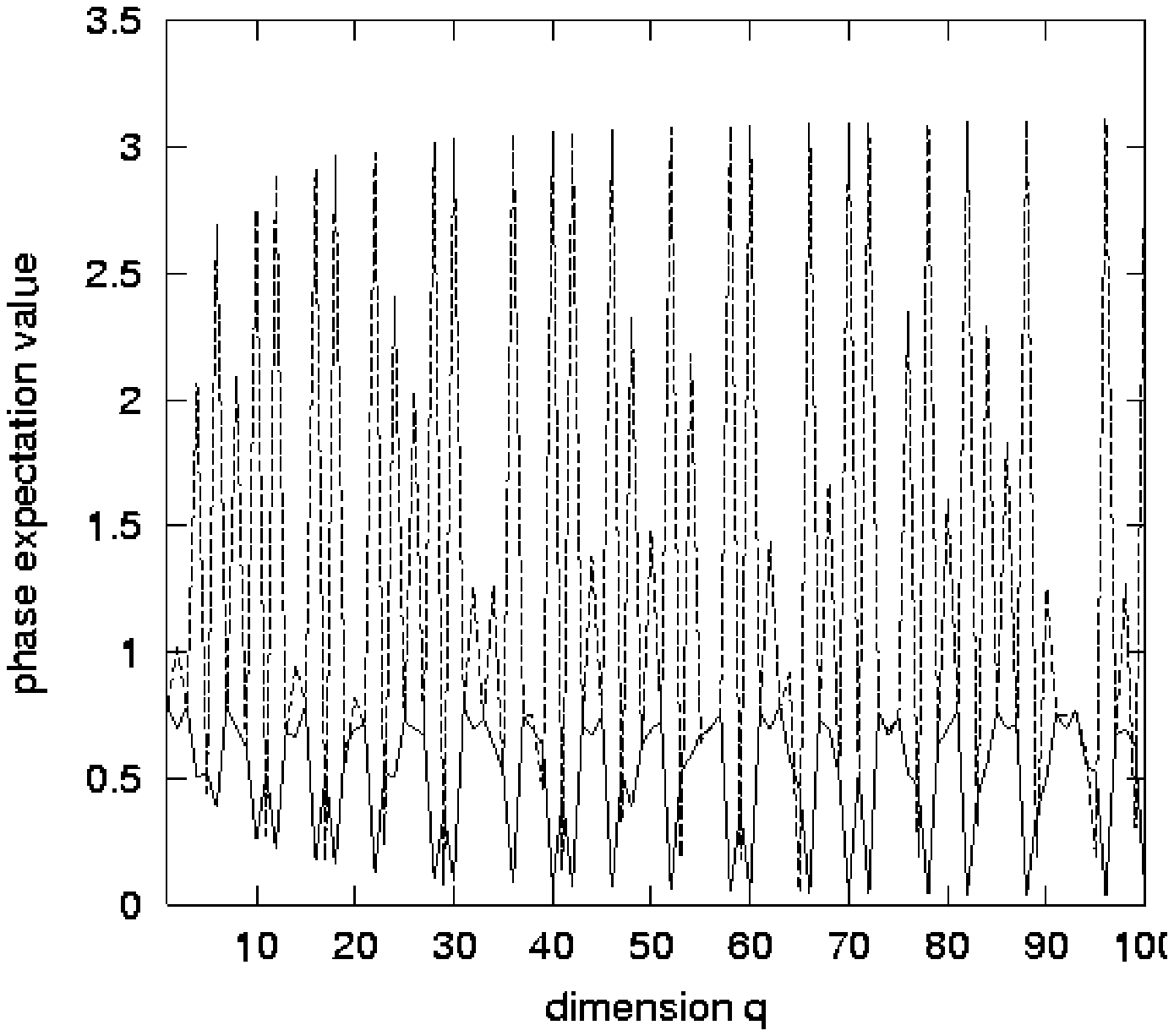}}
\vskip 3ex
\begin{center}
{\small{Fig. 1}\\
The phase expectation value versus the dimension $q$ of the
Hilbert space. Plain lines: $\beta=0$. Dotted lines: $\beta=1$
}
\end{center}
%
\vskip 1ex
\centerline{
\epsfxsize=220pt
\epsfbox{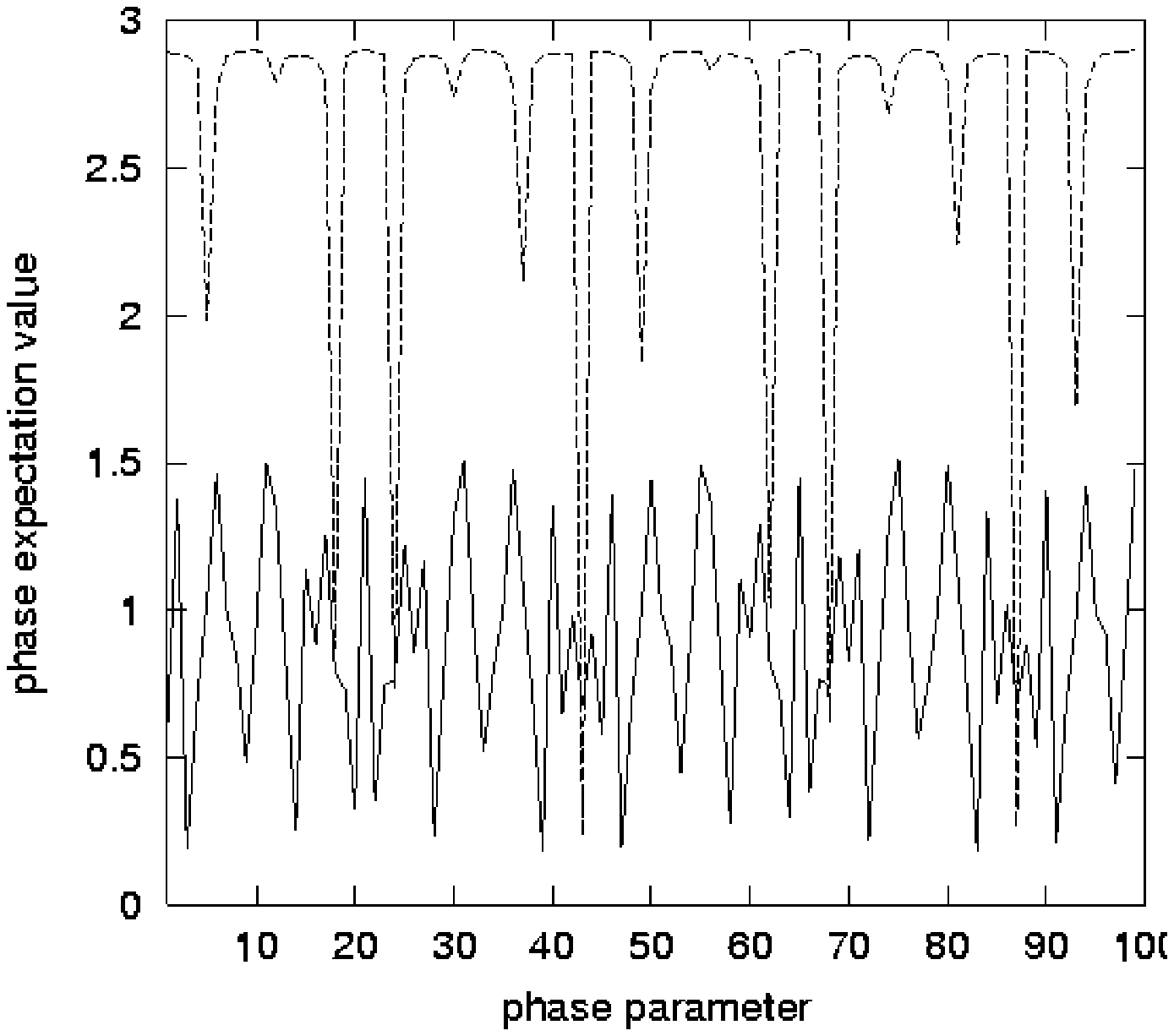}}
\vskip 3ex
\begin{center}
{\small{Fig. 2}\\
The phase expectation value versus the phase parameter
$\beta$. Plain lines: $q=15$. Dotted lines: $q=13$.
}
\end{center}
\vskip 1ex
\centerline{
\epsfxsize=220pt
\epsfbox{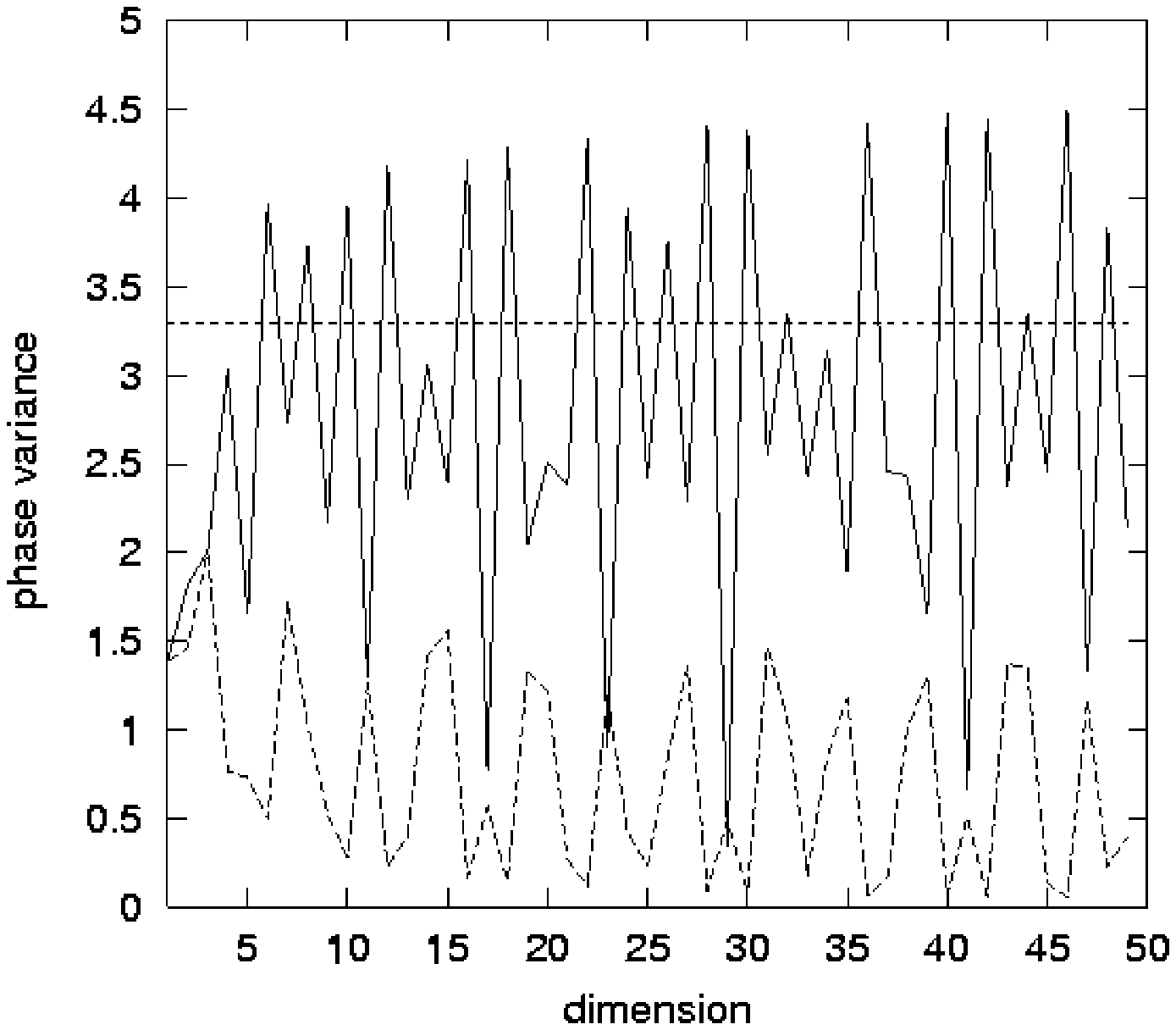}}
\vskip 3ex
\begin{center}
{\small{Fig. 3}\\
The phase variance versus the dimension q of the Hilbert
space. Plain lines: $\beta=1$. Dotted lines: $\beta=\pi$.
The horizontal line at $\pi ^2/3$ is the variance of the classical phase.}
\end{center}
%

\medskip

Rephrasing Pegg and Barnett, we consider a pure phase state $|f\rangle=\sum_{n=0}^{q-1}u_n|n\rangle$ having $u_n$ of the
form
\begin{equation}
u_n=\frac{1}{\sqrt{q}}\exp(i n\beta),
\end{equation}
where $\beta$ is a real phase parameter. Defining the phase
probability distribution as $\langle\theta'_p|f\rangle^2$; the phase
expectation value as $\langle \Theta_q^{{\rm lock}}\rangle=\sum_{p}\theta_p
\langle\theta'_p|f\rangle^2$; and the phase variance as ($\Delta
\Theta_q ^{2})^{{\rm lock}}=\sum_p (\theta_p-\langle \Theta_q^{{\rm lock}}\rangle)^2
\langle\theta'_p|f\rangle^2$, where the summations are taken from
$p=0$ to $q-1$ and $(p,q)=1$; the following result is obtained
\begin{eqnarray}
&\langle\Theta_q^{{\rm lock}}\rangle=\frac{\pi}{q^2}\sum_{n,l=0}^{q-1}
c_q(l-n) \exp[i\beta(n-l)],\\
&(\Delta \Theta_q^2)^{{\rm lock}}=4\langle \tilde{\Theta}_q^{{\rm lock}}\rangle
+\frac{\langle\Theta_q^{{\rm lock}}\rangle}{\pi}(\langle\Theta_q^{{\rm lock}}\rangle-2\pi),
\end{eqnarray}
with the modified expectation value
\begin{equation}
\langle\tilde{\Theta}_q^{{\rm lock}}\rangle=\frac{\pi}{q^2}\sum _{n,l=0}^{q-1}
\tilde{c}_q(l-n) \exp[i\beta(n-l)],\\
\end{equation}
 and the modified Ramanujan sums
 \begin{equation}
\tilde{c}_q(m)=\sum_{\stackrel{p=1}{(p,q)=1}}^q (p/q)^2 \exp\left(2i\pi
m\frac{p}{q}\right).
\end{equation}

Fig.~1 illustrates the phase expectation value versus the
dimension $q$ for two different values of the phase parameter
$\beta$. For $\beta=1$, there are peaks at dimensions $q=p^r$, which
are powers of a prime number $p$. For $\beta=0$, the peaks are
smoothed out due to the averaging over the Ramanujan sums.
Fig.~2 shows the phase expectation value versus the phase
parameter $\beta$ for two dimensions of the Hilbert space, $q=15$ and $q=13$, respectively. 
For the case of the prime number $q=13$, the phase expectation value is of
considerable higher level than for $q=15$, with absorption-like lines at isolated values of $\beta$. For
$q=15$, which is not a prime power, the
phase expectation value is at a much lower level and more random than for the case of the neighbour prime number.

Fig.~3 illustrates the phase variance versus the dimension $q$.
The case of $\beta=1$ also leads to peaks at prime powers. Similar to the
expectation value in Fig.~1, this case is thus reminiscent of the
Mangoldt function $\Lambda(n)$ defined as
$\ln p$ if $n$ is the power of a prime number $p$  and $0$
otherwise. This function arises in the frame of prime number theory
\cite{Planat1} in the formula for the logarithmic derivative of the Riemann zeta
function  
$-\frac{\zeta'(s)}{\zeta(s)}=\sum_{n=0}^{\infty}\frac{\Lambda(n)}{n^s}$.
Its mean value oscillates around $1$ with an error term which is
explicitely related to the positions of Riemann zeros of $\zeta(s)$ on the critical line
$s=\frac{1}{2}$. The error term shows a power spectral density
close to that of $1/f$ noise \cite{Planat1}. 
Finally, the phase variance is considerably smoothed
out for $\beta=\pi$ and is much lower than the classical limit
$\pi^2/3$. Thus, the parameter $\beta$ is useful in defining quantum phase-locked
states with small phase variances for a whole range of dimensions.

In conclusion, several interesting properties are pinpointed here as a result of 
introducing cyclotomic discreteness in the Pegg-Barnett quantum phase 
formalism. The interplay between results 
associated to prime number theory and quantum phase-locking phenomena looks
quite stimulating. We recall that the idea of quantum teleportation was initially 
formulated by Bennett {\em et al.} in finite-dimensional Hilbert space \cite{telep}.
Yet independently of this, one can conjecture that cyclotomic locking could play 
an important role in many fundamental tests of quantum mechanics related to 
quantum entanglement. As a matter of fact, Munro and Milburn \cite{mumi} have
already conjectured that the best way to see the quantum nature of correlations 
in entangled states is through the measurement of the observable, canonically 
conjugate of the photon number, i.e., the quantum phase. Moreover, in their paper 
dealing with the Greenberger-Horne-Zeilinger quantum correlations, Munro and Milburn presented
a homodyne scheme requiring discrete phase measurement, yet to be experimentally 
realized in the ultrahigh detector efficiency limit. Nevertheless, a homodyne scheme automatically
implies phase locking measurements with cyclotomy playing a role. 
We also recall that a universal algorithm for the optimal quantum state estimation of an arbitrary finite dimensional system 
in terms of a sequence of projectors
has been introduced by Derka {\em et al} \cite{derka}. In one of their two examples, the sequence of projectors is very similar to the Pegg-Barnett projectors
(in fact they noticed the analogy with the PB phase operator). Therefore, the quality of the cyclotomic phase estimation through a direct generalization
of the maximal {\em mean} fidelity introduced in their paper could be utilized as well. 
It may be possible that the type of quantum phase-locking described in
this paper may have some applications to the synchronization of
remote oscillators by quantum entanglement as already proposed in
\cite{Shariar}.

\end{document}